\documentstyle[12pt]{article}
\def\be{\begin{equation}}
\def\ee{\end{equation}}
\def\bea{\begin{eqnarray}}
\def\eea{\end{eqnarray}}

\begin{document}
\begin{flushright}
{BROWN-HET-1010} \\
{September 1995}
\end{flushright}
\vspace*{5mm}
\begin{center}
{\bf PHYSICS OF THE VERY EARLY UNIVERSE}\footnote{Report on the workshop
`Physics of the Very Early Universe' held at GR14, Aug. 6-12 1995, Florence,
Italy; to be published in the proceedings, ed. by M. Francaviglia et al (World
Scientific, Singapore, 1996).}\\
[10mm]

{ROBERT H. BRANDENBERGER}\\
[4mm]
Brown University Physics Department\\
Providence, RI  02912, USA\\
[10mm]

{\bf Abstract}\\
\end{center}

{In this report of the workshop on  `Physics of the Very Early Universe' held
at GR14, the recent status and open problems in a selected number of major
areas of interest are reviewed, focusing on attempts to develop a superstring
cosmology and on progress in understanding current theories of structure
formation.}

\section{Introduction}

`Physics of the Very Early Universe' has two main goals.  The first is to find
ways of using cosmological data to test and constrain theories of microphysics,
the second is to use microphysics theories to develop explanations for
observational facts which are unexplained in the context of the standard big
bang cosmology.

As an example of the first avenue of investigation, many grand unified theories
of particle physics can be ruled out since they predict an unacceptably large
flux of monopoles~\cite{1} or domain walls~\cite{2}, at least in the absence of
further assumptions about the theory such as the postulate of a period of
exponential expansion of space in the very early Universe.

Modern field-theory based scenarios of structure formation \cite{3} (such as
inflation-based cold dark matter and topological defect models) represent a
remarkable success of the second approach.  For the first time, causal theories
based on microphysics exist which can explain some aspects of the observed
highly correlated structure of inhomogeneities in our Universe and which make
many specific predictions which will be tested in future observations.

The workshop on `Physics of the Very Early Universe', held as part of GR14,
focused on both of the above-mentioned avenues of investigation.  A first
two-hour session was devoted to superstring cosmology and related topics, with
the main goal being to discover if unified theories of the fundamental forces
such as string theory make specific testable predictions for cosmology.  A
second three hour session dealt with recent developments concerning theories of
structure formation, in particular topological defect models and inflation.

\section{Superstring Cosmology and Related Topics}

Superstring theory \cite{4} remains the most promising quantum theory of all
fundamental forces including gravity.  Although certain ingredients of the
theory (such as the existence of supersymmetry in Nature) may be testable in
particle accelerator experiments, cosmology is the only arena in which many
string-specific aspects can be tested.  Superstring cosmology will most likely
become one of the most active areas of cosmology over the next decade.

Besides the graviton, the low energy spectrum of string theory includes the
dilaton, an antisymmetric tensor field, and in many superstring models other
low mass scalar fields, the moduli.  The stringy nature of the theory manifests
itself in the existence of winding modes, states of the string spectrum
corresponding to strings winding around the spatial manifold (most easily to
visualize in the case where space is toroidal).  String theory also admits many
new symmetries, e.g. $t$-duality \cite{5}, a symmetry which interchanges
momentum and winding modes.  It would be rather surprising if these new fields,
modes and symmetries would not lead to important effects in early Universe
cosmology.

The question which is currently receiving a lot of attention is whether string
theory in a natural way predicts inflation.  As is well known, the most simple
realizations of scalar field driven inflation are observationally ruled out
since they predict too large microwave anisotropies and density perturbations.
In order to obtain a particle-physics-based model of inflation which gives the
correct amplitude for the spectrum of density fluctuations, it is necessary to
use a scalar field potential with a very small number (e.g. a small coupling
constant or a mass scale much smaller than the Planck mass) built into it.
First attempts to study inflation in string theory were based on the effective
field theory limit of string theory and thus face the same problems as
conventional field-theory-based implementations. In fact, additional problems
arise, e.g. the moduli problem, the fact that new scalar modes called the
moduli are excited and come to dominate the energy density of the Universe
\cite{42}. Recently it has been realized that in string theory the dilaton can
induce a new type of inflation \cite{6}  (sometimes called `kinetic
inflation'), and in this context the usual problems of inflation might be
avoided.

The status of dilaton-driven inflation was reviewed at this meeting by
G.Veneziano \cite{7}.  The action for string-induced dilaton gravity is
\be
S = \int  d^4 x \, \sqrt{-g} \, e^{-\phi} \, \left[ R + \partial_{\mu} \phi
\partial^{\mu} \phi + V (\phi )\right] \, ,
\ee
with $\phi$ denoting the dilaton field.  At tree level and in the absence of
supersymmetry breaking, the dilaton potential $V(\phi)$ vanishes.  Note that
the above action is written in the string (or Brans-Dicke) frame, not the
Einstein Frame.  The above action has an important duality symmetry:  scale
factor duality.  If the backgrounds are only time dependent, and the dynamics
hence described by $\phi(t)$ and the scale factor $a(t)$, then
\bea
a & \rightarrow &\tilde{a} = a^{-1}\nonumber\\
\phi & \rightarrow &\tilde{\phi} = \phi - 2d \ln a
\eea
(where $d$ is the dimension of space) is a symmetry of the theory.  Combining
this symmetry with time reflection $t\rightarrow -t$, we can establish a
correspondence between an expanding decelerating solution and an expanding
accelerating solution in which the Hubble radius $H^{-1} (t)$ decreases as $t$
increases, called a pre-big-bang solution.

The pre-big-bang solution associated with the standard radiation-dominated
evolution is
\bea
a(t) & \sim &(-t)^{-1/2}\nonumber\\
\phi (t) & = &- 3 \ln (-t)
\eea
for $t<0$.  It is the rapid increase of the dilaton field which generates
inflation.  In the Einstein frame, the solution (3) corresponds to a model with
accelerated contraction (``super-inflationary").

If there were a smooth transition between the pre-big-bang cosmology (3) and
its self-dual FRW counterpart, this would lead to a very appealing cosmology
with no singularity and in which inflation (in the sense of $\ddot{a} > 0)$ can
be obtained without introducing a scalar potential $V(\phi )$.  It has,
however, been recently shown that no such smooth transition is possible at the
level of dilaton gravity \cite{8} (even allowing for a general potential
$V(\phi)$).  Similar problems arise when working with `other' moduli fields.
However, the transition between pre-big-bang and radiation dominated
cosmologies would occur at high curvatures, and hence the other string modes
will no longer be negligible.

R. Easther \cite{9} presented a numerical investigation of cosmological
solutions of the equations of motion obtained from a string-inspired action
which in addition to the dilaton also includes central charge and an
antisymmetric tensor field.  Various cosmological behaviors can be obtained,
but the fundamental problems which prevent the establishment of a
phenomenologically viable nonsingular cosmology cannot be overcome.

The hope has been expressed that dilaton gravity might be combined with the
limiting curvature construction of Ref. 10 in order to obtain a viable
nonsingular pre-big-bang type cosmology.  This limiting curvature construction,
reviewed in the Workshop by V. Mukhanov, is based on deriving a higher
derivative action for gravity in which by construction all solutions become
de-Sitter-like at high curvatures.  Higher energy modes of string theory might
be responsible for limiting the curvature in string cosmology.  Such modes
(when integrated out) are expected to give rise to higher derivative gravity
terms in the effective action.  Most higher derivative actions yield solutions
with worse singularity properties than Einstein gravity, and thus a higher
derivative action which manages to smooth out singularities must have a very
special form.  An action for which all homogeneous and isotropic solutions are
nonsingular
and approach de Sitter at high curvatures can be written in terms of
Lagrange multipliers $\phi_1$ and $\phi_2$ as~\cite{10}.
\be
S = \int d^4 x \sqrt{-y} \left[ R + \phi_1 R - \left( \phi_2 + \sqrt{3} \phi_1
\right) I_2^{1/2} + V_1 (\phi_1 ) + V_2 (\phi_2 )\right] \, ,
\ee
where $I_2$ is a curvature invariant with the property that (given metrics of
the specified symmetries) $I_2 = 0$ has de Sitter spaces as unique solutions.
In the case of cosmological metrics, a choice for $I_2$ is
\be
I_2 = \left( 4R_{\mu v} R^{\mu v} - R^2 \right) \, .
\ee
The potentials $V_1 (\phi_1 )$ and $V_2 (\phi_2 )$ are constructed to give the
correct Einstein limit for low curvatures (small values of $\phi_1$ and $\phi_2
$), to bound $R$ explicitly and to force $I_2$ to zero at large curvatures
(large values of $\phi_2$).

During the workshop, the hope was expressed that a similar solution to the
singularity problem could be found in string theory.  Note that the limiting
curvature construction cannot eliminate singularities which are present at all
times and other ``good" singularities (see Ref. 11 for a recent discussion of
the value of certain singularities) .  Note also that string theory may resolve
cosmological singularities in rather different ways (see e.g. Ref. 12).

String theory may also lead to an explanation for the number of detectable
spatial dimensions.  Recall that critical superstring theory is defined in
$9+1$ space-time dimensions.  Usually it is assumed that 6 spatial dimensions
compactify and become small.  Another point of view was advanced in Ref. 12.
There, it was suggested that, if the initial state of the Universe is a thermal
state with all spatial dimensions compactified at the string scale, then
winding modes which are present in thermal equilibrium and wrap around all of
the spatial dimensions will allow at most three spatial dimensions to expand to
macroscopic scales.  Winding modes in four or more large spatial dimensions do
not intersect, and hence cannot disappear.  The winding modes will keep six
dimensions at the string scale.  In three spatial dimensions, however, the
fundamental strings will evolve in a nontrivial way. Since the action for
fundamental strings is the same as the effective action which governs the
dynamics of cosmic strings,

it is possible to gain insight into the evolution of a system of fundamental
strings from cosmic string simulations.  Cosmic string evolution has been well
studied with the result that of the order 1 winding mode per Hubble volume
$H^{-3} (t)
$ will remain ($H$ being the Hubble expansion constant), giving rise to a
subdominant contribution to the energy density and not preventing expansion.

The intuition expressed above that winding modes will prevent space from
expanding is Newtonian.  Applying Einstein's equations would yield the result
that winding increases the expansion rate.  However, the Einstein equations do
not respect $t$-duality of string theory. In Ref. 13 it has been shown that if
the effects of the dilaton are taken into account, both $t$-duality and the
Newtonian intuition that winding prevents expansion are restored, lending
support to the speculations of Ref. 12.

Further support for the above speculations were presented during the Workshop
by M. Sakellariadou~\cite{14}.  She presented results concerning string
propagation in 3,4 and 5 spatial dimensions, demonstrating that in greater than
3 dimensions the decay time for long strings (curvature radius greater than
$H^{-1}$ ) is much larger than in 3 dimensions.  The results were based on
numerical simulations using a code which solves the string equations of motion
exactly on a lattice in flat space-time and which can be extended to a
(nonexact) expanding Universe code~\cite{15}.  Various thermal properties of
strings and the equilibration process for the string network were studied, as
well as its dependence on space dimensionality.

A speculative solution of the singularity problem was presented by M. Mohazzab
in a poster contribution~\cite{16}.  The idea is that space is made up of fixed
size elementary building blocks and that the expansion of the Universe
corresponds to un-crumpling of the assembly of blocks.  This scenario involves
a time dependent effective dimension of space.

Unified microscopic theories can also be constrained by comparing the predicted
density inhomogeneities and microwave anisotropies with observations.  For a
special class of Kaluza-Klein cosmologies, this analysis was done in work
presented by V. Faraoni~\cite{17}.

\section{Cosmic String and Structure Formation}

As indicated in the Introduction, the development of theories of structure
formation based on causal microphysical laws has been a major achievement of
modern cosmology.  It is now very likely that a large number of quantitative
observations can be explained in a causal physical manner.

Two mechanisms have been discussed in depth. The first is based on quantum
fluctuations originating and increasing in wavelength during a period of
inflation in the very early Universe. The second is based on topological
defects (in particular cosmic strings) which are assumed to form during a phase
transition taking place in the very early Universe~\cite{3}.

Inflation-based models were invoked in the plenary lectures of J. Primack and
M. Turner.  However, all inflation-based models suffer -- in the absence of a
new and more appealing realization of the scenario such as what could emerge
from superstring cosmology -- from a small number problem:  without adjusting
parameters in the inflaton potential to very small values (as quantified in a
model-independent way in Ref. 18), the theory predicts too large fluctuations
and resulting cosmic microwave (CMB) anisotropies.  The scale of symmetry
breaking of the phase transition producing topological defects, in contrast, is
a scale at which there already is evidence (based on the convergence of the
standard model coupling constants which can be deduced from accelerator
experiment data) that new physics occurs, namely
\be
\eta \simeq 10^{16} GeV \, .
\ee

Several presentations during the Workshop were devoted to recent progress in
the cosmic string theory of structure formation.  T. Kibble gave an overview of
the model and discussed a new analysis of the formation of defects~\cite{19}.
For first order phase transitions, the initial defect density is determined by
the bubble nucleation rate. For a second order transition, it was long believed
that the mean separation of defects is given by the correlation length at the
Ginsburg temperature, a time slightly later than when the critical temperature
is reached. However, the new analysis presented by Kibble demonstrates that (in
agreement with an earlier analysis by Zurek~\cite{20}) dynamical considerations
determine the time at which the domain structure freezes in. The new research
has confirmed (in agreement with the findings of Ref. 21) that there is no
difference between the formation probability of defects in local and global
theories.  In no circumstance (barring inflation) is the causality bound
violated

(at least one defect per horizon volume).  In a related talk A. Melfo discussed
the geometrical suppression factor of the defect formation probability in a
first-order phase transition stemming from the fact that at the time of bubble
collision, not all of the bubbles have the same size~\cite{22}.

In talks by R. Caldwell~\cite{23} and by P. Shellard and R. Battye~\cite{24},
new results on cosmological consequences of cosmic strings were presented.  R.
Caldwell showed preliminary results of the first computation of the spectrum of
CMB anisotropies in the cosmic string theory based on numerical simulations of
the cosmic string network, but considering for the moment only the Sacks-Wolfe
effect.  P. Shellard and R. Battye have studied the effects of back reaction on
the spectrum of gravitational waves predicted by local strings.  The main
effect is to decrease the amplitude of the spectrum on wavelengths probed by
the millisecond pulsar, and thus slightly relaxing the pulsar constraint on
strings.

The results presented at this Workshop confirm that the cosmic string model is
a promising model of structure formation.  It predicts (like inflation) a scale
invariant primordial spectrum of density perturbations
\be
P(k) \sim k \, ,
\ee
where
\be
P (k) = < \vert {\delta \rho (k) \over \rho_0} \vert^2 > \, .
\ee
However, the density field $\rho (x)$ is not predicted to be a random field.
Long, rapidly moving strings will lead to a sheet-like structure of the
distribution of galaxies on scales larger than the Hubble radius at the time of
equal matter and radiation $t_{eq}$ when matter perturbations can start to
grow~\cite{25}.  The normalizations of the model from CMB anisotropy
measurements and large-scale structure date (e.g. large-scale peculiar
velocities) are consistent~\cite{26}.  The specific predictions of the cosmic
string model will be tested by the new deeper redshift surveys which are in
preparation.

One further advantage of the cosmic string model of structure formation is that
the dark matter of the Universe can be hot~\cite{27} (e.g. a 30eV tau
neutrino).  Within the context of inflationary models, the dark matter must
consist --at least in a spatially flat Universe-- to more than 70\% of cold
dark matter~\cite{28}.  A model with cosmic strings and hot dark matter also
solves another problem of CDM inflationary models: too much power predicted on
small scales compared to observations when the model is normalized to fit the
COBE results~\cite{29}.  If cosmic strings seed structure, then accretion on
small scales is delayed but not prevented by free streaming.

The specific signature of cosmic strings in the CMB is the presence of line
discontinuities produced by lensing of light induced by the conical structure
of space perpendicular to a string~\cite{30}.  These discontinuities, however,
are only identifiable in CMB maps with an arc-minute beam width~\cite{31}.  The
overall spectrum of CMB anisotropies on large scales is scale invariant as in
inflation-based models.  An outstanding issue is a careful calculation of the
acoustic contribution to degree scale anisotropies.  For some recent work in
this direction see e.g. Ref. 32.

Unfortunately, the initial string distribution is not yet well enough
established to allow for precise determination of many of the predictions of
the model.  This, at the moment, is the most serious problem for the string
model.

\section{Other Topics}

This section is a brief summary of some of the other contributed talks in this
Workshop covering a fairly wide range of subjects.  The discussion given here
can only be very superficial and is intended rather to advertise some recent
results rather than to give a representative scientific discussion.

There has been a lot of recent interest in black hole solutions in models of
gravity such as dilaton gravity (Brans-Dicke theory) in which the gravitational
constant $G$ depends on time.  It is well known that severe constraints on the
number density of black holes formed in the early Universe can be derived by
demanding that the $\gamma$ rays emitted from evaporating black holes do not
exceed the observed $\gamma$ ray background.  B. Carr~\cite{33} presented new
results generalizing these constraints to black holes in Brans-Dicke theories
and other models with time-dependent $G$ (i.e. with gravitational memory).

R. Gregory presented the results of new work~\cite{34} addressing the black
hole no-hair conjecture.  Specifically, she showed analytical and numerical
evidence for the existence of solutions to the coupled Einstein-Abelian Higgs
Equations of motion representing a Schwarzschild black hole threaded by a U(1)
cosmic string.  This analysis provides more evidence that the no-hair
conjecture is false provided that matter fields are allowed to be present at
the event-horizon.  It was also shown that cosmic string solutions exist for
which the strings terminate in black holes.

In another talk, L. Ford~\cite{35} discussed light cone fluctuations and
primordial gravitons.  It has been shown that a stochastic background of
gravitons produces fluctuations of the light cone.  This leads to two
consequences.  First, light propagation becomes a statistical phenomenon with
some photons appearing to travel faster and others slower than the speed of
light.  Second, these fluctuations remove ultraviolet divergences in some but
not all quantum field expectation values.

Several talks were devoted to issues in the inflationary Universe scenario.  A.
Barvinsky presented the results of an investigation of the initial conditions
for inflation from the point of view of quantum cosmology~\cite{36}.  The
``no-boundary" proposal for the wave function was applied to a model based on a
non-minimally coupled scalar field.

The ``fluctuation problem"~\cite{18} of the inflationary Universe scenario was
the topic of a talk by M. Morikawa~\cite{37}.  He proposed to take fluctuation
and dissipation effects into account in the computations of generation and
classicalization of density perturbations.  The claim was made that particle
production caused by cosmic expansion leads to a friction term in the equation
for classical perturbations.  The end result is that the predicted amplitude of
density fluctuations at late times (after reheating) is smaller than what is
predicted by the standard calculations, thus lessening the ``small number
problem" for inflation mentioned at the beginning of Section 3 of this Report.
Similar results were obtained in Ref. 38.  It will be very important to analyze
this issue carefully in view of its relevance for the inflationary Universe
scenario.

The talk by D. Wands~\cite{39} was also related to the issue of fluctuations in
inflationary cosmology.  He presented an analysis of the constraints on the
parameters of general scalar-tensor theories of gravity stemming from demanding
that perturbations produced during inflation not exceed the cosmological
constraints from recent CMB observations.

Two final talks were by N. Sakai~\cite{40} and by R. Mann~\cite{41}.  Sakai
discussed fluctuations of the gravitational constant induced by primordial
bubbles in models of extended inflation and constraints which can be derived
from these considerations.  Mann discussed an attempt to describe elementary
particles in terms of a classical model and why these attempts have failed.

\section{Conclusions}

This workshop was a lively one and demonstrated the vitality of this area of
research.  Investigations concerning `Physics of the Early Universe' are driven
by the continuing progress in observational cosmology.  New data, in particular
concerning CMB anisotropies and the large-scale structure of the Universe, is
becoming available at a steady pace, fueling research into theories of
structure formation and leading to tighter constraints on particle physics
models.  On the other hand, many conceptual issues remain to be resolved, e.g.
the connection between string theory and cosmology, to mention just one of the
topics discussed at this meeting.

I wish to thank all the speakers for coming to Florence and sharing their new
research results.  The local organizing committee of GR14 deserves a round of
applause for their great efforts in handling this large meeting.


\end{document}